\documentclass[useAMS,a4paper,fleqn,usenatbib]{mnras}
\usepackage{color}
\usepackage{graphicx}	% Including figure files
\usepackage{amsmath}	% Advanced maths commands
\usepackage{amssymb}	% Extra maths symbols
\usepackage{multicol}        % Multi-column entries in tables
\usepackage{bm}		% Bold maths symbols, including upright Greek
\usepackage{pdflscape}	% Landscape pages

\usepackage{multirow}

\usepackage[T1]{fontenc}
\usepackage{ae,aecompl}

\usepackage{txfonts}
\newcommand{\ngc}{NGC~612}
\newcommand{\fc}{4C~73.08}
\newcommand{\tc}{3C~452}

\newcommand{\xmm}{{\em XMM-Newton}}
\newcommand{\nus}{{\em NuSTAR}}
\newcommand{\chandra}{{\em Chandra}}
\newcommand{\suz}{{\em Suzaku}}
\newcommand{\integral}{{\em INTEGRAL}}
\newcommand{\sax}{{\em BeppoSAX}}
\newcommand{\swift}{{\em Swift}}

\newcommand{\aer}[3]{$#1^{+ #2}_{- #3}$}
\newcommand{\aerm}[3]{#1^{+ #2}_{- #3}}

\newcommand{\ser}[2]{$#1 \pm #2$}
\newcommand{\serm}[2]{#1 \pm #2}
\newcommand{\serexp}[3]{$#1 \pm #2 \times 10^{#3}$}
\newcommand{\lsup}[1]{$< #1$}

\newcommand{\expo}[2]{$ #1 \times 10^{#2}$}
\newcommand{\tento}[1]{$10^{#1}$}
\newcommand{\tentom}[1]{10^{#1}}
\newcommand{\expom}[2]{ #1 \times 10^{#2}}

\newcommand{\fluxcgs}{erg~cm$^{-2}$~s$^{-1}$}
\newcommand{\lumcgs}{erg~s$^{-1}$}

\newcommand{\sqcm}{cm$^{-2}$}

\newcommand{\chisq}{\chi^{2}}
\newcommand{\rchisq}{\chi^{2}/\textrm{dof}}
\newcommand{\dchi}{\Delta \chi^{2}}
\newcommand{\ddof}{\Delta \textrm{dof}}

\newcommand{\nh}{N_{\textrm{H}}}

\newcommand{\sigmat}{\sigma_{\textrm{T}}}
\newcommand{\lowflux}{F_{\textrm{2--10keV}}}
\newcommand{\highflux}{F_{\textrm{10--100keV}}}
\newcommand{\lowlum}{L_{\textrm{2--10keV}}^{\textrm{intr}}}
\newcommand{\highlum}{L_{\textrm{10--100keV}}^{\textrm{intr}}}
\newcommand{\nhi}{N_{\textrm{H}\,\textsc{i}}}

\newcommand{\fek}{Fe~K~$\alpha$}

\newcommand{\oiii}{O\,{\sc iii}}

\newcommand{\hi}{H\,{\sc i}}

\newcommand{\xspec}{{\sc xspec}}

\newcommand{\nthcomp}{{\sc nthcomp}}
\newcommand{\mytorus}{{\sc mytorus}}

\newcommand{\zphabs}{{\sc zphabs}}
\newcommand{\apec}{\textsc{apec}}
\newcommand{\pexrav}{\textsc{pexrav}}

\usepackage[space]{grffile}
\usepackage[normalem]{ulem}
\usepackage{comment}
\bibpunct{(}{)}{;}{a}{}{,}

\defcitealias{panessa2016}{P16}
\newcommand{\panessa}{\citetalias{panessa2016}}

\begin{document}
\title[Where are Compton-thick radio galaxies?]{Where are Compton-thick radio galaxies? A hard X-ray view of three candidates}

\author[F. Ursini]
	{F. Ursini,$^{1}$
		\thanks{e-mail: \href{mailto:ursini@iasfbo.inaf.it}{\texttt{ursini@iasfbo.inaf.it}}} 
	L. Bassani,$^{1}$
	F. Panessa,$^{2}$
	A. Bazzano,$^{2}$
	A.~J. Bird,$^{3}$
	A. Malizia,$^{1}$
	\newauthor	
	and
	P. Ubertini$^{2}$
%	and
%	friends
%	P.-O. Petrucci,$^{1,2}$
%	G. Matt,$^{3}$ 
%	S. Bianchi,$^{3}$ 
%	M. Cappi,$^{4}$ \newauthor
%	B. De Marco,$^{5}$ 
%	A. De Rosa,$^{6}$
%	J.~Malzac,$^{7,8}$ 
%	A.~Marinucci,$^{3}$
%	G. Ponti$^{5}$ 
%	and
%	A. Tortosa$^{3}$ 
\\
	$^1$ INAF-IASF Bologna, Via Gobetti 101, I-40129 Bologna, Italy. \\
	%\and
	$^2$ INAF/Istituto di Astrofisica e Planetologia Spaziali, via Fosso del Cavaliere, 00133 Roma, Italy.\\
	$^3$ School of Physics and Astronomy, University of Southampton, SO17 1BJ, UK.
}

\date{Released Xxxx Xxxxx XX}

%\pagerange{\pageref{firstpage}--\pageref{lastpage}} \pubyear{2014}

\maketitle

\label{firstpage}

\begin{abstract}
We present a broad-band X-ray spectral analysis of the radio-loud active galactic nuclei \ngc, \fc\ and \tc, exploiting archival data from \nus, \xmm, \swift\ and \integral. These Compton-thick candidates are the most absorbed sources among the hard X-ray selected radio galaxies studied in \cite{panessa2016}. We find an X-ray absorbing column density in every case below \expo{1.5}{24} \sqcm, and no evidence for a strong reflection continuum or iron K $\alpha$ line. Therefore, none of these sources is properly Compton-thick. We review other Compton-thick radio galaxies reported in the literature, arguing that we currently lack strong evidences for heavily absorbed radio-loud AGNs. 
\end{abstract}

\begin{keywords}
	galaxies: active -- galaxies: Seyfert -- X-rays: galaxies -- X-rays: individual: NGC~612, 4C~73.08, 3C~452
\end{keywords}

\section{Introduction}\label{sec:intro}
A significant fraction of active
galactic nuclei (AGNs) are known to be intrinsically obscured by gas and dust surrounding their central engine. 
%In the soft X-ray band, the photoelectric absorption produces a low-energy cut-off depending on the column density $\nh$ of the absorber. 
According to the unified model, the presence of a torus at pc scales can explain the properties of obscured AGNs and the dichotomy with unabsorbed sources, simply due to a different orientation with respect to the line of sight \cite[e.g.][]{antonucci&miller,antonucci1993}.

Obscured AGNs are considered an important class of sources for two main reasons. First of all, most AGNs in the local universe are absorbed \cite[$\nh > \tentom{22}$ \sqcm; e.g.][]{risaliti1999,bassani1999}. As shown by different X-ray surveys, the observed fraction of absorbed AGNs generally ranges between 40 and 60 per cent \cite[e.g.][and references therein]{burlon2011}. However, their \textit{intrinsic} fraction, i.e. corrected for selection biases, is around 80 per cent \cite[e.g.][]{malizia2012}. Moreover, obscured AGNs are an important ingredient of both the cosmic X-ray background \cite[e.g.][]{fabian&iwa99,gilli2007} and of the infrared background, where most of the absorbed high-energy radiation is re-emitted \cite[e.g.][]{reg2002}.

In particular, Seyfert 2  galaxies in which the absorbing material has $\nh \geq \sigmat^{-1} = 1.5 \times 10^{24}$ \sqcm, where $\sigmat$ is the Thomson cross-section, are called Compton-thick (CT). Such high column densities produce a photoelectric cut-off at energies of 10 keV or more. Hard X-rays data are thus needed to properly study CT sources \cite[see, e.g., the analysis of \sax\ data by][]{matt2000}. A number of local CT AGNs have been detected thanks to recent hard X-ray surveys with \integral\ \cite[][]{sazonov2008,malizia2009,malizia2012} and \swift\ \cite[e.g.][]{burlon2011,ricci2015}. At high redshift, the rest-frame hard X-ray emission becomes observable below 10 keV and CT AGNs have been detected using deep surveys by \xmm\ \cite[e.g.][]{lanzuisi2015} and \chandra\ \cite[e.g.][]{brightman2014}. While the observed fraction of CT AGNs is generally found to be of a few percent, their intrinsic fraction could be as high as 20--30 per cent. This estimate remains highly uncertain, owing to the difficulties in detecting and classifying such heavily obscured sources.

It is a matter of debate whether differences exist in the absorption properties between different classes of AGNs, depending on their physical parameters \cite[like the accretion rate or the luminosity, e.g.][]{elitzur2006}. For example, the fraction of obscured AGN is observed to decrease with increasing X-ray luminosity \cite[e.g.][]{ueda2003,lafranca2005,sazonov2015}, although with some uncertainties related to luminous but heavily obscured sources \cite[]{mateos2017}, and possibly with decreasing redshift \cite[e.g.][]{lafranca2005}. Radio galaxies are particularly interesting in this respect, because the presence of a jet could have an impact on the circumnuclear environment and on the obscuring medium. Large fractions of X-ray obscured radio galaxies are generally found \cite[e.g.][]{sambruna1999,evans2006,hardcastle2009,wilkes2013}. In particular, \cite{wilkes2013} reported a fraction of 20 per cent of CT sources in a high-redshift sample ($1 < z < 2$). However, only a few radio galaxies in the local Universe have been reported as CT: 
Mrk~ 668 \cite[][]{guainazzi2004}, 
3C~284 \cite[][]{hardcastle2006}, 
3C~223 \cite[][]{lamassa2014,corral2014},
3C~321 \cite[]{evans2008,severgnini2012},
%Swift~J0601.9-8636 \cite[][]{ueda2007,eguchi2009}, 
\tc\ \cite[][]{fioretti2013},
TXS~2021+614 \cite[]{siemiginowska2016} and PKS~1607+26 \cite[]{tengstrand2009,siemiginowska2016}.

In a recent work, \citet[][P16 hereafter]{panessa2016} investigated the role of absorption in radio galaxies. The sample analysed by \panessa\ consisted of 64 AGNs showing extended radio emission, selected from hard X-ray catalogues by \cite{bassani2016}. Being hard X-ray selected, this sample is relatively unbiased towards absorption, at least up to a few $\times \tentom{24}$ \sqcm. \panessa\ derived the column density distribution, and found an observed fraction of absorbed sources of 40 per cent (75 per cent among the type 2s). 
However, the fraction of CT sources was found to be no more than 2--3 per cent, whereas 7--10 per cent would have been expected \cite[e.g.][]{malizia2012}.

In this paper, we focus on the most absorbed radio galaxies reported in \panessa, with the aim of studying their  broad-band X-ray spectra and to well constrain their column density. We take advantage of archival \nus\ data, which are very helpful in constraining the spectral shape of absorbed sources thanks to the high-energy coverage \cite[e.g.][]{balokovic2014}. The structure of the paper is as follows. In Section \ref{sec:obs} we present the sources and their archival X-ray data used. In Section \ref{sec:analysis}, we present the spectral analysis. We discuss the results in Section \ref{sec:discussion}, and summarize the conclusions in Section \ref{sec:summary}.

\section{Observations and data reduction}\label{sec:obs}
%\subsection{Sources selection}
We selected the two sources that, according to the results of \panessa, show an $\nh$ consistent with being larger than $\tentom{24}$ \sqcm. These are \ngc\ and \fc\ (VII Zw 292), for which \panessa\ reported $\log \nh = \serm{24.02}{0.2}$ and $\aerm{23.96}{0.20}{0.16}$, respectively. We also included \tc\ ($\log \nh = \aerm{23.77}{0.06}{0.07}$ according to \panessa) which has been labelled as a CT candidate by \cite{fioretti2013}. 

\subsection{The sources}
\begin{table}
		\begin{center}
			%\scriptsize
			\caption{Basic data of the three sources selected from \panessa. \label{tab:sources}}
			\begin{tabular}{ l c c c } 
				\hline \hline  
				& \ngc & \fc & \tc   \\ \hline
				Optical classification & Sy2 & Sy2  & Sy2  \\
				Radio morphology & FRI/II & FRII & FRII  \\
				Host galaxy type & S0 & E &  E \\
				Redshift & 0.029 771&0.058 100&0.081 100 \\
				%			Luminosity distance &&& \\
				Galactic $\nh$ (\tento{20} \sqcm) &1.85&2.41&11.9 \\ \hline
				
			\end{tabular}
		\end{center}
	\end{table}
The basic data of \ngc, \fc\ and \tc\ are reported in Table \ref{tab:sources}.

\ngc\ is a rare case of a radio source hosted by a spiral galaxy \cite[e.g.][]{veron2001}, with a prominent dust lane along the disc \cite[e.g.][]{ekers1978,veron2001}. The optical spectrum is consistent with the presence of a reddened young stellar population of age $<0.1$ Gyr \cite[]{holt2007}. The infrared properties indicate an active star formation taking place in a warped disc \cite[]{duah2016}. The radio morphology is also peculiar, as it shows an eastern radio lobe of type FRII and a western lobe of type FRI \cite[]{gopal2000}. Furthermore, \ngc\ is surrounded by a 140 kpc-wide disc of neutral hydrogen, and an \hi\ bridge is seen towards the neighbour galaxy NGC 619 \cite[]{emonts2008}. The AGN in \ngc\ is optically classified as a Seyfert 2 \cite[e.g.][]{veron2006,parisi2009} and it is absorbed in the X-rays.
Using \suz\ data, \cite{eguchi2009} found $\nh \simeq \expom{1.1}{24}$ \sqcm\ and suggested a scenario where the X-ray source is obscured by a torus with a half-opening angle of 60-70 deg, seen from a nearly edge-on angle.

\fc\ is a giant radio galaxy, with a linear size of around 1 Mpc \cite[e.g.][]{bassani2016}. It exhibits a complex morphology, but overall consistent with a FRII \cite[e.g.][]{lara2001,strom2013}. The AGN is optically classified as a Seyfert 2 \cite[e.g.][]{hewitt1991,parisi2014}. This source is poorly studied in the X-rays, however \cite{evans2008} inferred an absorbing column density of around \expo{9}{23} \sqcm\ for the nucleus, from the analysis of \xmm\ data.

\tc\ is a radio galaxy with a symmetrical FRII morphology \cite[e.g.][]{black1992}, optically classified as a Seyfert 2 \cite[e.g.][]{veron2006}. 
From \chandra\ data, \cite{isobe2002} found the nucleus to be absorbed by a column density of around \expo{6}{23} \sqcm, and detected a diffuse X-ray emission from the lobes. Similar results were obtained by \cite{evans2006} with the same data set. Using \xmm\ data, \cite{shelton2011} performed a detailed X-ray study of both the AGN and the lobes, constraining their inverse Compton emission and environmental impact. No significant variations were found from the comparison with the \chandra\ observation, while the absorbing column density was found to be around \expo{4}{23} \sqcm\ for the AGN. The same column density was reported by \cite{fioretti2013}, who analysed \suz\ data. Finally, according to the results of \cite{fioretti2013}, the hard energy band of the X-ray spectrum is dominated by Compton reflection off cold matter, with a reflection fraction \cite[as defined in][]{pexrav} found to be $R>400$. This unphysical value could be due to an inhomogeneous circumnuclear material, where a significant fraction of the solid angle is covered by a gas thicker than that along the line of sight \cite[]{fioretti2013}.

\subsection{The X-ray data}
For the three sources we analysed archival \nus\ data. Being sensitive in the 3--79 keV band, \nus\ \cite[][]{harrison2013nustar} allows us to constrain both the spectral shape at high energies and the absorbing column density. We also included the average 70-month \swift/BAT spectra \cite[][]{bat70}\footnote{\url{https://swift.gsfc.nasa.gov/results/bs70mon/}} to extend the coverage up to 195 keV, and \integral/IBIS data \cite[]{ibis} for \tc. Moreover, to better constrain the column density, we included the soft X-ray spectra by \xmm\ \cite[][]{xmm} when possible, namely in the case of \ngc\ and \fc.

We report in Table \ref{tab:log} the logs of the archival \nus\ data sets of the three sources.
We reduced the \nus\ data using the standard pipeline (\textsc{nupipeline}) in the \nus\ Data Analysis Software (\textsc{nustardas}, v1.8.0; part of the \textsc{heasoft} distribution as of version 6.22), using calibration files from \nus\ {\sc caldb} v20170817. We extracted the spectra 
%and light curves 
using the standard tool {\sc nuproducts} for each of the two hard X-ray detectors, which reside in the corresponding focal plane modules A and B (FPMA and FPMB). We extracted the source data from circular regions with a radius of 75 arcsec, and the background from a blank area close to the source. The spectra were binned to have a signal-to-noise ratio larger than 3 in each spectral channel, and not to oversample the instrumental resolution by a factor greater than 2.5. The spectra from the two detectors were analysed jointly, but not combined.

The \xmm\ data were processed using the \xmm\ Science Analysis System (\textsc{sas} v16). We used the EPIC-pn data, because of the much larger effective area of pn than that of the MOS detectors. The source extraction radii and screening for high-background intervals were determined through an iterative process that maximizes the signal-to-noise ratio \cite[see][]{pico2004}. We extracted the background from circular regions with a radius of 50 arcsec, while the source extraction radii were in the range 20--40 arcsec. We binned the spectra to have at least 30 counts per spectral bin, and not oversampling the instrumental resolution by a factor larger than 3.

The \integral\ spectrum of \tc\ consists of ISGRI data from several pointings between revolution 12 and 530 \cite[from the fourth IBIS catalogue;][]{bird2010}. The data extraction was carried out following the procedure described in \cite{molina2013}.

\begin{table}
	\begin{center}
		%\scriptsize
		\caption{Logs of the \textit{NuSTAR} observations of the sources. \label{tab:log}}
		\begin{tabular}{ c c c c } 
		\hline	\hline Source &  Obs. Id. & Start time (UTC)  & Net exp.\\ 
			& & yyyy-mm-dd & (ks)  \\ \hline 
			\ngc 
			%& \xmm  & 0312190201  & 2006-06-26 & 10 \\ 
			%& \nus  
			& 60061014002 & 2012-09-14 & 17 \\ \hline 
			\fc 
			%\xmm & & &   \\
			%& \nus 
			& 61060374002 & 2016-12-05 & 13 \\ \hline 
			\tc 
			%& \xmm  &   0552580201  & 2008-11-30 & \\
			%& \nus 
			& 60261004002 & 2017-05-01 & 52 \\ \hline
			%&& 11200110003 & 2014-07-11 & 800 & 290\\ \hline
		\end{tabular}
	\end{center}
\end{table} 

\section{Spectral analysis}\label{sec:analysis}
The spectral analysis was carried out with the \xspec\ 12.9.1 package \cite[][]{arnaud1996}, using the $\chisq$ minimisation technique. 
All errors are quoted at the 90 per cent confidence level.
As we explain below, we fitted the 3--79 keV \nus\ spectra and the 14--195 keV \swift/BAT spectra simultaneously, leaving the cross-calibration constants free to vary, after checking for consistency. In the case of \ngc\ and \fc, we included \xmm/pn data to extend the broad-band analysis down to 0.3 keV. We did not include \xmm\ data of \tc, because the pn spectrum was not consistent with the \nus\ one (see Sect. \ref{subsec:452}). Finally, in the case of \tc\ we exploited \integral/IBIS data.

In all cases, we first fitted the data with a phenomenological model including an absorbed power law. Then, we tested a more physical absorption model, i.e. a gas torus \cite[\mytorus:][]{mytorus}. We always included Galactic absorption, fixing the hydrogen column densities to the values obtained from the \hi\ map of \citet[as given by the tool \textsc{nh} in the \textsc{HEASoft} package]{kalberla2005}. For all models, we adopted the chemical abundances of \cite{angr} and the photoelectric absorption cross-sections of \cite{vern}.

\begin{figure}
	\includegraphics[width=\columnwidth]{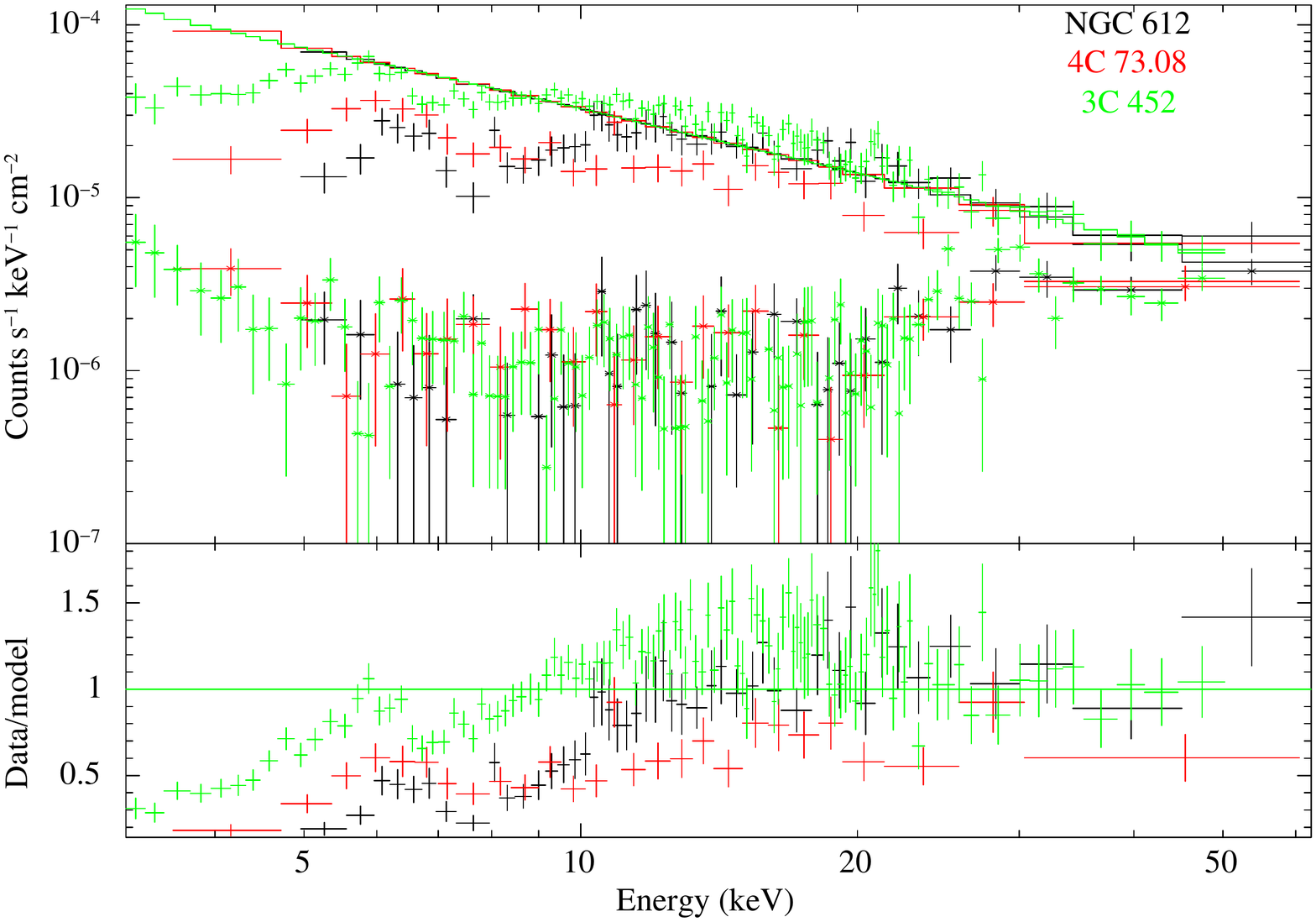}
	\caption{Upper panel: \nus/FPMA spectra of \ngc\ (black), \fc\ (red) and \tc\ (green) plotted with the corresponding background level. Lower panel: residuals when fitting simultaneously the spectra above 10 keV with a simple power law. The data were binned for plotting purposes.
		\label{fig:spectra}}
\end{figure}

In Fig. \ref{fig:spectra}, we plot for comparison the \nus\ spectra of the three sources with the best-fitting power law above 10 keV (with tied parameters), showing the effects of absorption at lower energies.
\subsection{\ngc}
\nus\ observed this source in 2012 with a net exposure of 17 ks, as part of the \nus\ serendipitous survey \cite[]{nus_serendip}. \xmm\ observed the source in 2006 (Obs. Id. 0312190201), with a net exposure of 10 ks. Before using the \xmm/pn data, we checked the flux variation of the source between the \xmm\ and \nus\ observations. We thus fitted the \nus\ spectrum with a simple power law modified by photoelectric absorption (\zphabs\ model in \xspec) to derive the 3--10 keV flux, which was found to be $\expom{\serm{1.15}{0.08}}{-12}$ \fluxcgs. The photon index was $\Gamma=\serm{1.60}{0.16}$ and the column density $\nh= \expom{\serm{84}{12}}{22}$ \sqcm. Next, we fitted the \xmm/pn spectrum in the 3--10 keV band with the same model with fixed parameters, only allowing for flux variability. The 3--10 keV flux measured by pn was $\expom{\serm{1.66}{0.11}}{-12}$ \fluxcgs. The flux variation is thus of 30--50 per cent. However, given the large error bars and the cross-calibration uncertainties of 10 per cent between \nus\ and \xmm/pn \cite[]{crosscal}, we did not consider it a dramatic variation and we fitted the two spectra simultaneously, allowing for a scaling factor between pn and \nus. We verified a posteriori that the spectral parameters are consistent with those measured by \nus\ alone (see Table \ref{tab:fits}). 
We also included the 70-month \swift/BAT spectrum, which is consistent with \nus\ with a cross-normalization constant of \ser{1.25}{0.13}.

As a first step, we fitted the \nus, \xmm/pn and \swift/BAT spectra in the 0.3--195 keV range, with a phenomenological model including a power law multiplied by \zphabs\ (``\zphabs\ model''). To find a good fit to the pn data, we needed to include further emission components that were not absorbed by \zphabs, i.e. not originating from the central X-ray source. This ``soft excess'' above the absorbed power law is ubiquitous in Seyfert 2s \cite[e.g.][]{turner1997,gmp2005}, and can be due to scattering of the primary continuum in an optically thin region \cite[e.g.][]{turner1997,ueda2007} and/or photoionization of circumnuclear gas \cite[e.g.][]{gb2007}. For \ngc, we included: (i) a secondary power law, with the parameters tied to those of the primary one, to model optically-thin scattered continuum; we added a free multiplicative constant corresponding to the scattered fraction $f_s$, expecting $f_s$ to be of the order of a few percent or less \cite[e.g.][]{turner1997}; (ii) a thermal component (\apec\ in \xspec), to model photoionized emission. We allowed for free cross-normalization constants between \nus/FPMA and FPMB ($K_{A-B}$), pn ($K_{A-pn}$) and BAT ($K_{A-BAT}$). In \xspec\ notation, the model reads: \textsc{const*phabs*(zphabs*powerlaw + powerlaw + apec)}.

The results are summarized in Table \ref{tab:fits}. We find a good fit ($\rchisq=131/132$) and constraints on the photon index and column density.
We tested the presence of an exponential cut-off at high energies, but this parameter was unconstrained. We also tested the presence of a \fek\ emission line at 6.4 keV, which is a common feature in the X-ray spectra of obscured AGNs \cite[e.g.][]{gmp2005}, and which can be particularly prominent in CT sources dominated by reflection \cite[e.g.][]{matt1996}. Including a narrow Gaussian line, we found a rest-frame energy of \ser{6.31}{0.08} keV and a line flux of \serexp{2.6}{1.6}{-6} photons \sqcm\ s$^{-1}$. The fit is improved, albeit not dramatically ($\rchisq = 124/130$, i.e. $\dchi/\ddof = -7/-2$, with a probability of chance improvement of 0.03 calculated with an $F$-test). The equivalent width of the line is \aer{110}{80}{60} eV.
Then, we tested the presence of a Compton-reflected component by replacing the power law with the \pexrav\ model \cite[]{pexrav}. This model includes the continuum reflected off an infinite slab of infinite optical depth, which can be considered an approximation for the reflection off a finite-density torus. We fixed the inclination angle at 60 deg, and we assumed solar abundances. However, the fit is not significantly improved, and the reflection fraction is poorly constrained (we only derive an upper limit of 1.6).

Next, we tested a more physical absorption model by a torus. We replaced the \zphabs\ component with the \mytorus\ model, which describes absorption with self-consistent Compton reflection and iron fluorescent lines from a gas torus with a half-opening angle of 60 deg \cite[][]{mytorus,mytorus2}. In this case, we had to use pn data above 0.5 keV, because this is the lower bound for which \mytorus\ is valid.
The inclination angle of the torus was fixed at 90 deg, with no improvement by leaving it free to vary. We included the scattered and line emission components, 
linking their column densities to that of the absorber (``coupled'' model). The photon indexes and normalizations of the scattered and line components were tied to those of the primary power law, allowing for a free relative normalization by means of a multiplicative constant\footnote{The fluxes (or equivalent widths) of the fluorescence lines are not free parameters in \mytorus, because they are determined by the incident continuum (for a detailed explanation, see Chapter 7 of the \mytorus\ manual.}. 
We assumed the standard \mytorus\ configuration $A_S=A_L$, where $A_S$ is the scaling factor for the scattered component and $A_L$ that of the line component \cite[][]{mytorus2}. 
In \xspec\ notation, the model reads: \textsc{const1*phabs*(mytorusZ*powerlaw + apec + powerlaw +  const2*mytorusS + const2*mytorusL)}.

\begin{figure}
	\includegraphics[width=\columnwidth,trim={0.5cm 0 2cm 0},clip]{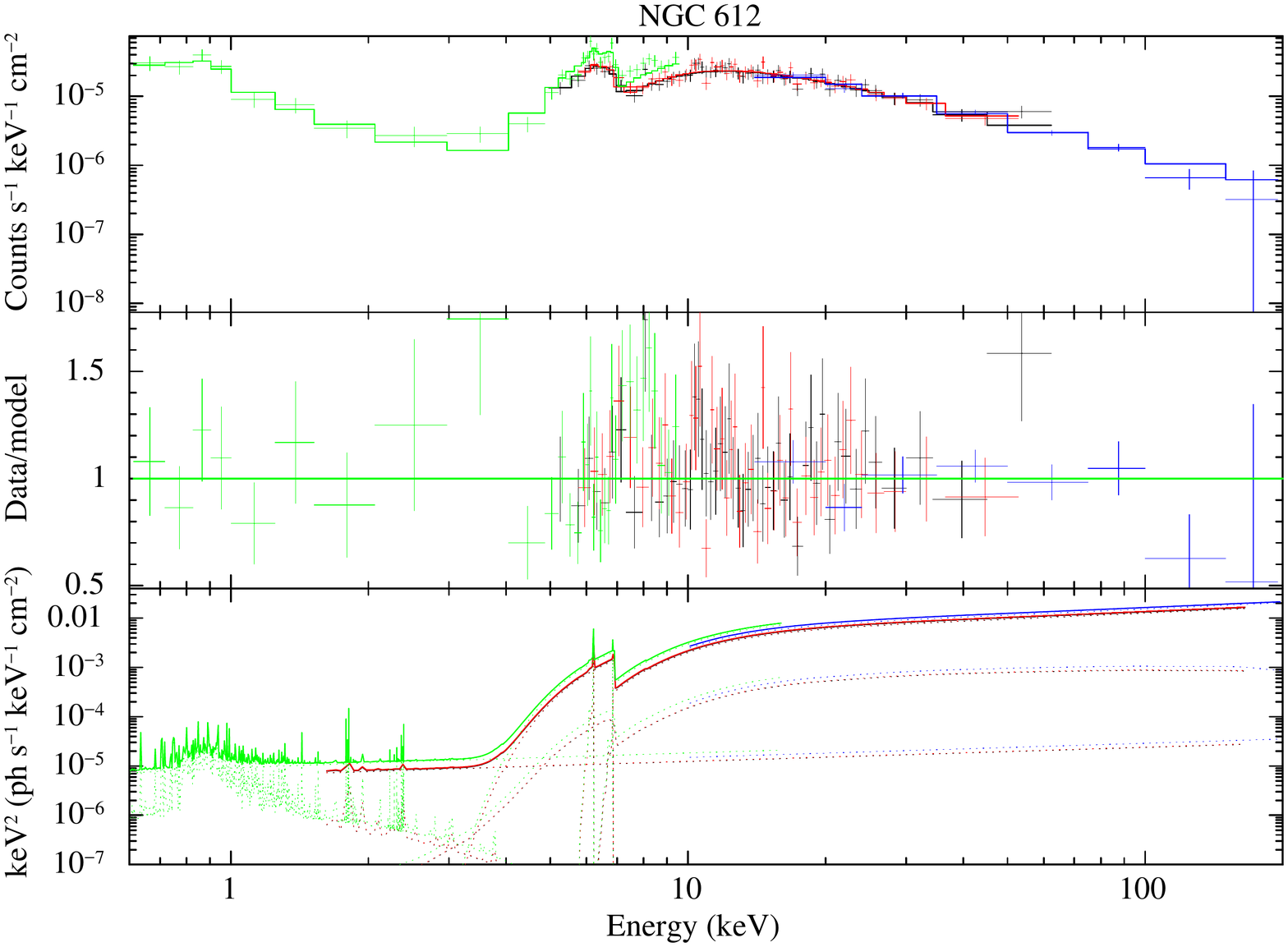}
	\caption{Broad-band X-ray data and best-fitting \mytorus\ model for \ngc\ (see Table \ref{tab:fits}). Upper panel: spectra and folded model from \nus/FPMA (black), FPMB (red), \xmm/pn (green), \swift/BAT (blue). Middle panel: ratio of data to model. Lower panel: best-fitting model $E^2 f(E)$, consisting of the absorbed primary power law, the scattered power law, the \mytorus\ reflection component and the thermal component \apec.
		\label{fig:ngc612}}
\end{figure}

\begin{figure}
	\includegraphics[width=\columnwidth,trim={1cm 0 2cm 0},clip]{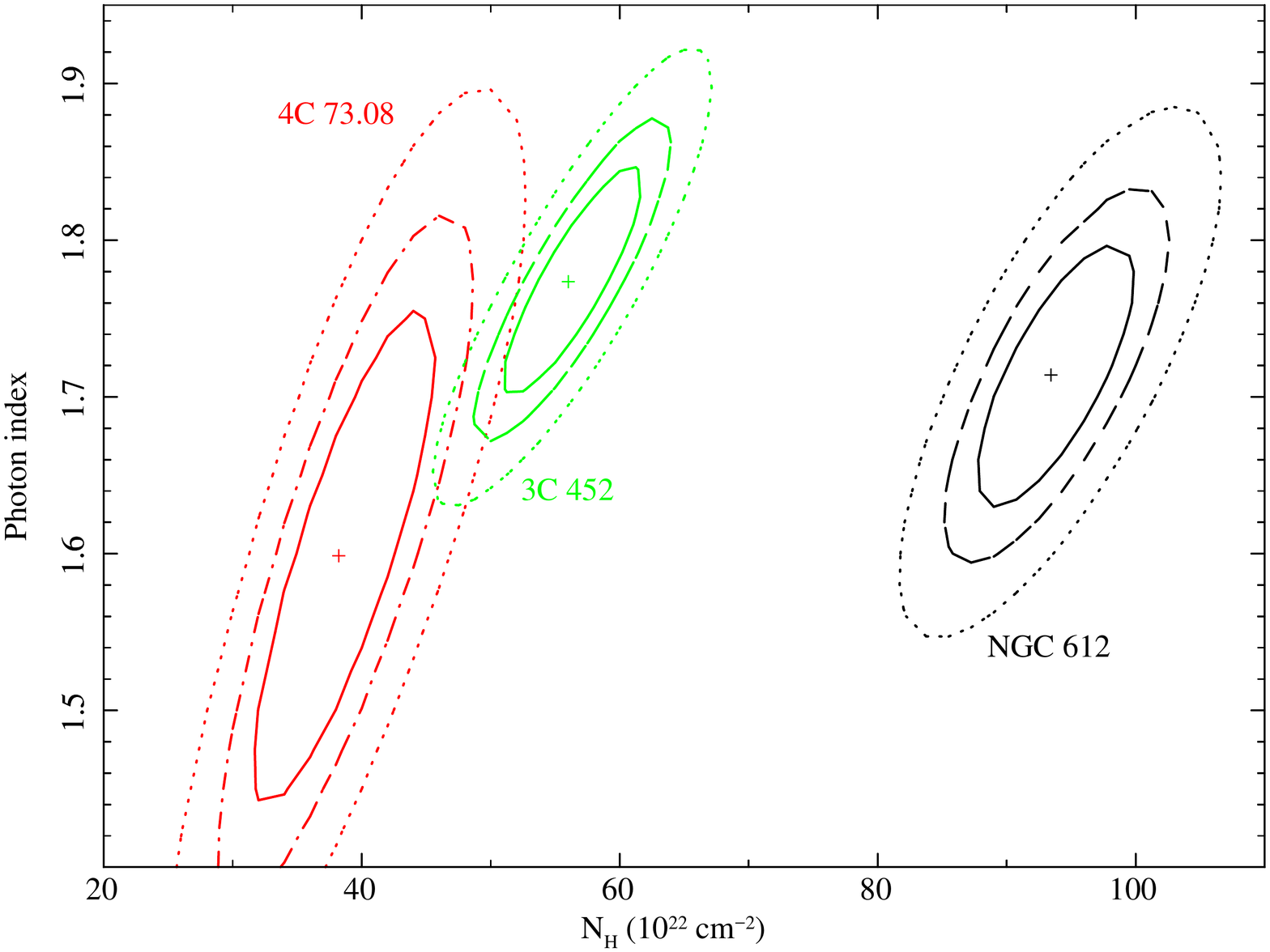}
	\caption{Contour plots of the column density $\nh$ versus the photon index of the primary power law (\mytorus\ model, see Table \ref{tab:fits}) for \ngc\ (black), 
		\fc\ (red) and \tc\ (green). Solid, dashed and dotted lines correspond to 68 per cent, 90 per cent and 99 per cent confidence levels, respectively.
		\label{fig:cont_myt}}
\end{figure}

The results for the \mytorus\ model are reported in Table \ref{tab:fits} (see also the contours in Fig. \ref{fig:cont_myt}), while in Fig. \ref{fig:ngc612} we show the data, residuals and best-fitting model. The column density is $\expom{\serm{94}{7}}{22}$ \sqcm. The parameters are essentially consistent with those of the \zphabs\ model, and the fit is slightly improved ($\rchisq = 121/129$ i.e. $\dchi/\ddof=-10/-3$, but we have 2 dof less because of the different energy range). The parameter $A_S$ is \aer{0.25}{0.22}{0.19}, i.e. relatively low, indicating a weak Compton-scattered continuum. To further test the significance of this result, we fixed $A_S$ to unity, finding a worse fit ($\rchisq = 142/130$ i.e. $\dchi/\ddof=+21/+1$). We note that $A_S=1$ would correspond to the default, steady-state configuration of \mytorus, in which the torus has a covering factor of 0.5 \cite[]{mytorus}. However, $A_S$ cannot be immediately related to the covering factor, because it is a simple scalar, while the precise shape of the scattered continuum depends on several parameters, such as the geometry of the system, elemental abundances and putative time delays between the scattered and intrinsic continuum \cite[e.g.][]{mytorus2}. Anyway, the case $A_S=0$ would correspond to a geometry in which the torus subtends a small solid angle, so that the scattered continuum and the fluorescence emission are negligible. In this case, the \mytorus\ model simply consists in photoelectric absorption plus additional attenuation due to Compton scattering. This last effect is not included in models like \zphabs, and this is the reason why the normalization of the primary power law, as well as the estimated intrinsic luminosity, is larger using \mytorus.  

\subsection{\fc}
This source was observed by \nus\ in 2016 with a net exposure of 13 ks, and by \xmm\ in 2007 (Obs. Id. 0404050601) with a net exposure of 7.5 ks. As for \ngc, we first fitted the \nus\ data with a power law modified by \zphabs, finding a 3--10 keV flux of \serexp{1.5}{0.1}{-12} \fluxcgs. The photon index was found to be \ser{1.4}{0.2}, and the column density \serexp{40}{10}{22} \sqcm. Applying this model to the \xmm/pn data in the 3--10 keV band, we found a flux of \serexp{1.05}{0.15}{-12} \fluxcgs. The flux variation is thus of around 50 per cent. Nevertheless, we fitted the pn data jointly with \nus, allowing for a free scaling factor and verifying that the spectral parameters were consistent with those measured by \nus. 
Finally, we included the 70-month \swift/BAT spectrum, finding a cross-normalization constant of \ser{0.76}{0.20} with respect to \nus.

Like for \ngc, we first fitted the \nus, \xmm/pn and \swift/BAT spectra in the 0.3--195 keV range using the \zphabs\ model. In the case of \fc, however, we fitted the ``soft excess'' above the absorbed power law with a secondary power law only, as a thermal component was not needed. The results are summarized in Table \ref{tab:fits}. The fit is statistically good ($\rchisq=145/152$) and, despite the larger error bars compared with \ngc, the column density is constrained and below \tento{24} \sqcm.
We found no constraint on the high-energy cut-off, and a hint of a narrow \fek\ emission line with rest-frame energy of \ser{6.45}{0.15} keV, line flux of \serexp{4}{3}{-6}  photons \sqcm\ s$^{-1}$ and equivalent width of \ser{120}{100} eV ($\rchisq = 142/150$, i.e. $\dchi/\ddof = -3/-2$, with a probability of chance improvement of 0.2). Then, we tested the presence of a Compton-reflected component with \pexrav, finding no major improvement and a poor constraint on the reflection fraction (the upper limit was 2.2). 

\begin{figure}
	\includegraphics[width=\columnwidth,trim={0.5cm 0 2cm 0},clip]{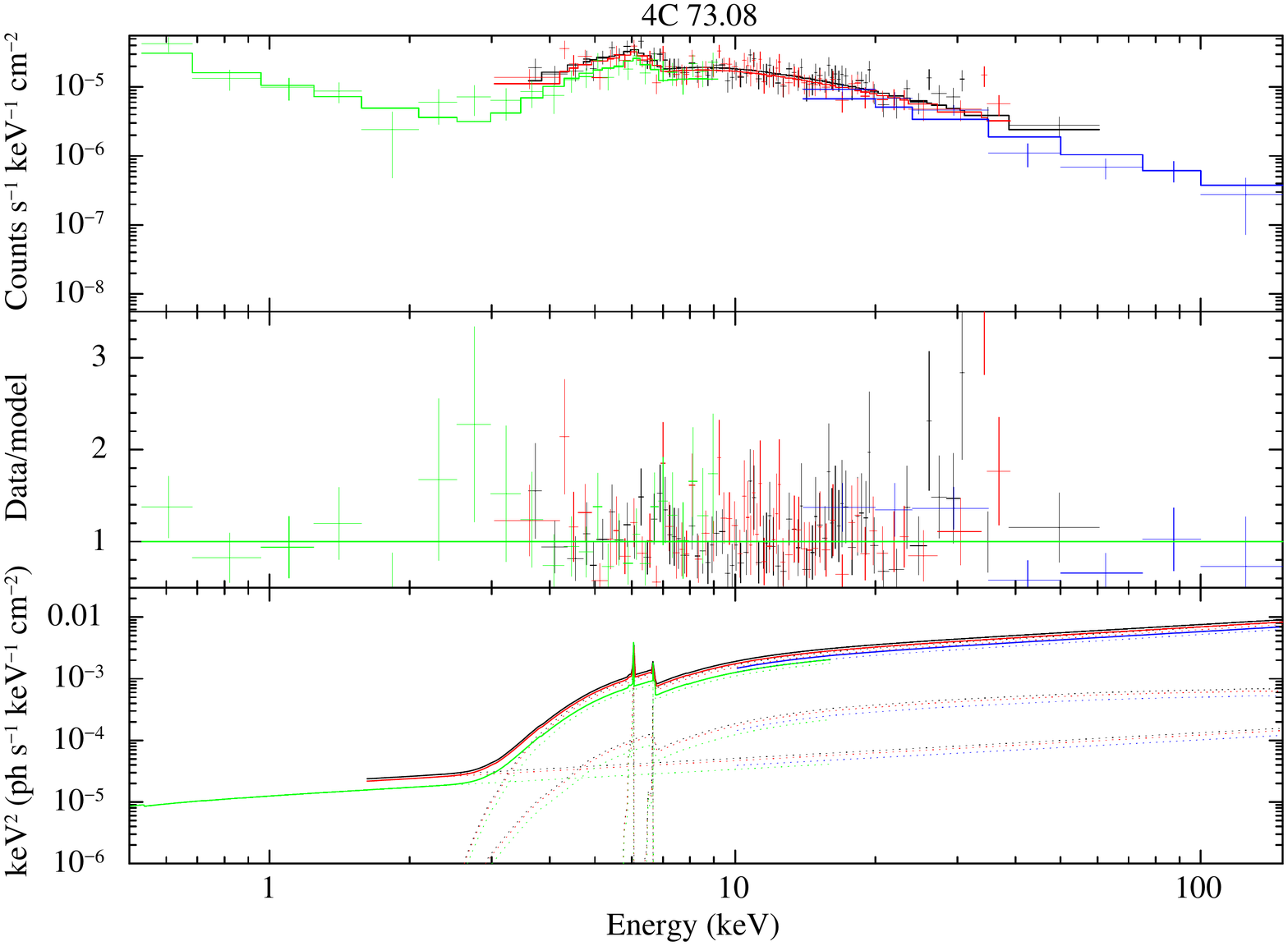}
	\caption{Broad-band X-ray data and best-fitting \mytorus\ model for \fc. Upper panel: spectra and folded model from \nus/FPMA (black), FPMB (red), \xmm/pn (green), \swift/BAT (blue). Middle panel: ratio of data to model. Lower panel: best-fitting model $E^2 f(E)$, consisting of the absorbed primary power law, the scattered component and the \mytorus\ reflection component (which, however, is consistent with zero). 
		\label{fig:4c73}}
\end{figure}

Finally, we tested the \mytorus\ model (see Table \ref{tab:fits} and the contours in Fig. \ref{fig:cont_myt}) using the same standard setting as for \ngc. The best-fitting parameters are consistent within the errors with those of the \zphabs\ model, and the fit is slightly improved ($\rchisq = 140/151$ i.e. $\dchi/\ddof=-5/-1$). In particular, the column density is $\expom{\serm{39}{8}}{22}$ \sqcm. The parameter $A_S$ is not well constrained, as we find only an upper limit of 2.4. The data, residuals and best-fitting model are shown in Fig. \ref{fig:4c73}. 

\subsection{\tc}\label{subsec:452}
\nus\ observed this source in 2017 with a net exposure of 52 ks, as part of the \nus\ extragalactic survey. We fitted the \nus\ spectrum jointly with the 70-month \swift/BAT spectrum, which had a cross-normalization constant of \ser{0.9}{0.1} with respect to \nus. \tc\ was also observed by \integral/IBIS with an exposure of 100 ks, and is part of the fourth \integral/IBIS catalogue \cite[]{bird2010,malizia2012}.
We thus included the IBIS spectrum in the 20--150 keV range, finding a cross-normalization constant of \ser{1.3}{0.5} with respect to \nus.
In the case of \tc, we did not use soft X-ray data.
%The \nus+\swift/BAT+\integral\ data set was sufficient to yield good constraints on both the photon index and the column density.
Although the source was observed by \xmm\ in 2008 and by \suz\ in 2007, the corresponding soft X-ray spectra are not fully consistent with \nus. Indeed, for \nus\ we found a photon index of \ser{1.76}{0.07}, while \cite{shelton2011} reported a photon index of \aer{1.26}{0.32}{0.27} for \xmm/pn, and \cite{fioretti2013} reported \aer{1.55}{0.14}{0.11} for \suz. However, the \nus+\swift/BAT+\integral\ data set was sufficient to yield good constraints on both the photon index and the column density.

Also in this case, we first fitted the spectra with a power law modified by \zphabs. However, the fit was relatively poor ($\rchisq=276/243$) with significant residuals in the 3-5 keV band. We found a much better fit ($\rchisq=220/242$, i.e. $\dchi/\ddof=-56/-1$) by including a secondary power law, with a scattered fraction $f_s = \serm{0.09}{0.01}$. Part of this excess could be actually due to extended emission related to the radio structure \cite[]{isobe2002,fioretti2013}. However, this contamination is relatively small above 3 keV and we do not expect it to affect the measurement of the column density.
The best-fitting parameters are reported in Table \ref{tab:fits}.
We did not find any statistical improvement of the fit by adding a narrow \fek\ line at 6.4 keV (the equivalent width being less than 50 eV), nor by adding a reflection component with \pexrav, with an upper limit on the reflection fraction of 0.8. Including an exponential high-energy cut-off results in a minor improvement of the fit ($\rchisq=215/241$, i.e. $\dchi/\ddof=-5/-1$ and a probability of chance improvement of 0.02). Formally, the cut-off energy was found to be \aer{120}{360}{55} keV within the 90 per cent confidence level. To further explore this possibility, we replaced the simple power law with a thermal Comptonization model, namely \nthcomp\ \cite[][]{nthcomp1,nthcomp2}. We assumed illumination from a disc blackbody with a seed photon temperature of 10 eV. We found a nearly equivalent fit ($\rchisq=218/241$; $\dchi=+3$), with $\Gamma= \serm{1.75}{0.06}$ and only a lower limit on the electron temperature of 30 keV.

\begin{figure}
	\includegraphics[width=\columnwidth,trim={0.5cm 0 2cm 0},clip]{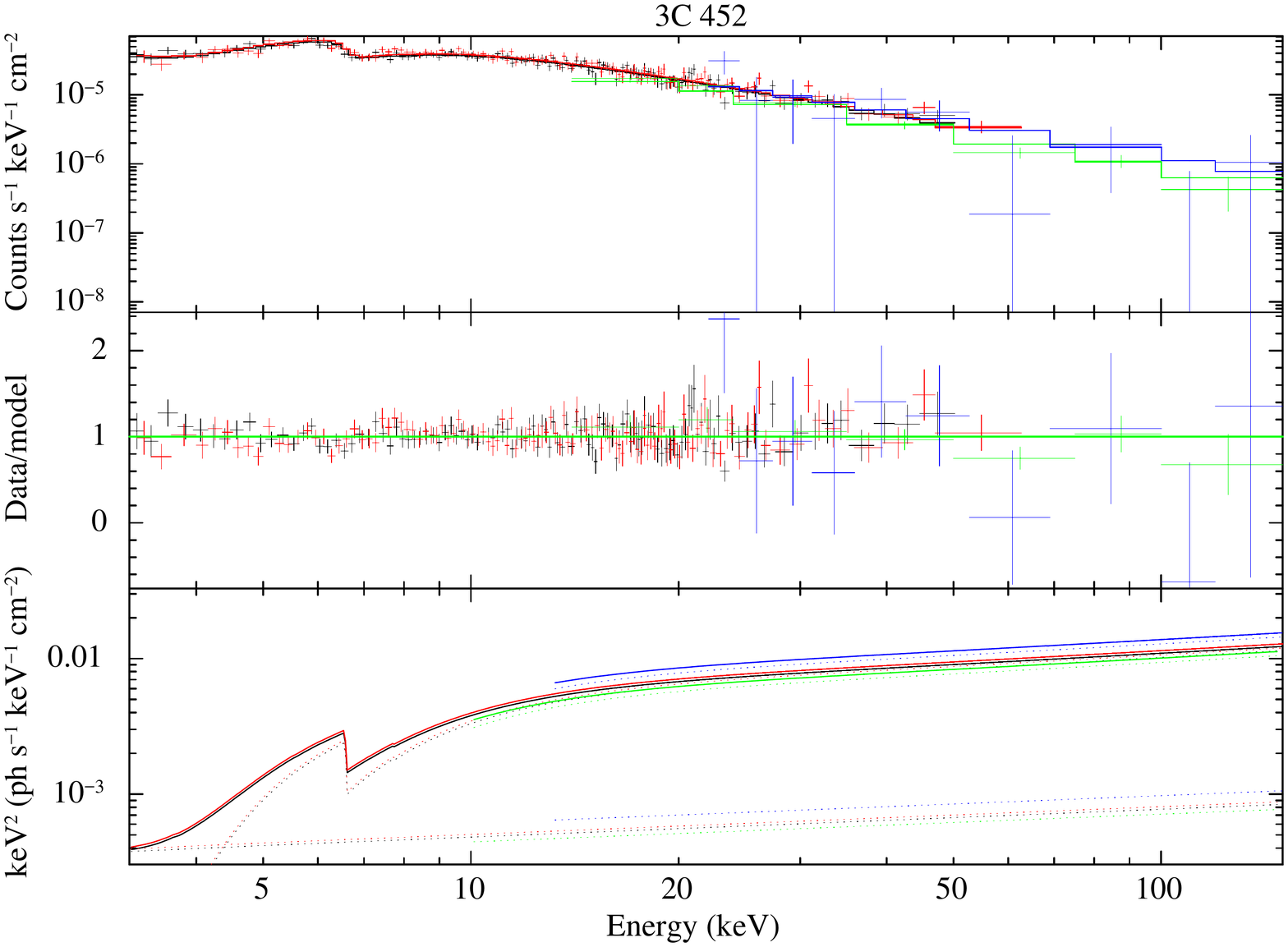}
	\caption{Broad-band X-ray data and best-fitting model for \tc. Upper panel: spectra and folded model from \nus/FPMA (black), FPMB (red), \swift/BAT (green), \integral/IBIS (blue). Middle panel: ratio of data to model. Lower panel: best-fitting model $E^2 f(E)$, consisting of the absorbed primary power law and the scattered component. Only the absorption component of \mytorus\ is needed, while the reflection component is not significant.
		\label{fig:3c452}}
\end{figure}

Then, we tested the \mytorus\ model with the standard setting (see Table \ref{tab:fits} and Fig. \ref{fig:cont_myt}). In this case, we found no improvement relative to the initial \zphabs\ fit ($\rchisq=223/241$, i.e. $\dchi/\ddof=+3/-1$). The column density is $\expom{\serm{57}{6}}{22}$ \sqcm, and the parameter $A_S$ is less than 0.2 (fixing $A_S$ to unity gives a worse fit with $\rchisq=250/242$, i.e. $\dchi/\ddof=+27/+1$). The data, residuals and best-fitting model are shown in Fig. \ref{fig:3c452}.

\begin{table}
	\begin{center}
		%\scriptsize
		\caption{Best-fitting parameters of the baseline model (\zphabs\ model) and of the \mytorus\ model. \label{tab:fits}}
		\begin{tabular}{ l c c c} 
			\hline \hline
		& \ngc\ & \fc & \tc  \\ \hline 
			\multicolumn{4}{c}{\zphabs\ model}  \\ \hline 
		$\Gamma$ & \aer{1.64}{0.09}{0.19}& \ser{1.61}{0.17}& \ser{1.76}{0.07}\\ 
		$N_{\textsc{pow}}$ (\tento{-3}) &\ser{2.7}{1.0}&\aer{1.2}{0.8}{0.5}&\aer{3.7}{1.0}{0.8}\\
		$\nh$ (\tento{22} \sqcm) &\aer{93}{5}{8}&\ser{40}{8}&\ser{57}{6}\\
		$f_s$ (\tento{-2})&\aer{0.27}{0.15}{0.08}&\ser{2}{1}&\ser{9}{1}\\
		$kT_{\apec}$ (keV)&\aer{0.86}{0.15}{0.19}&-&-\\
		$N_{\apec}$ (\tento{-6})&\aer{3.2}{1.8}{1.4}&-&-\\
		$\lowflux$ (\tento{-12} \fluxcgs)&\ser{1.11}{0.07}&\ser{1.6}{0.1}&\ser{3.37}{0.09}\\
		$\highflux$ (\tento{-11} \fluxcgs)&\ser{3.0}{0.2}&\ser{1.6}{0.2}&\aer{2.9}{0.1}{0.2}\\
		$\lowlum$ (\tento{43} \lumcgs)&\ser{2.3}{0.3}&\ser{4.2}{1.0}&\ser{19}{2}\\
		$\highlum$ (\tento{43} \fluxcgs)&\ser{6.6}{0.5}& \ser{13}{2}&\ser{44}{2}\\
		$K_{A-B}$ & \ser{1.02}{0.07}&\ser{0.92}{0.08}&\ser{1.04}{0.03}\\
		$K_{A-pn}$ & \ser{1.50}{0.15}&\ser{0.67}{0.11}&-\\
		$K_{A-BAT}$ & \ser{1.25}{0.13}&\ser{0.76}{0.20}&\ser{0.9}{0.1}\\
		$K_{A-IBIS}$ & -&-&\ser{1.3}{0.5}\\
		$\rchisq$&$131/132$&$145/152$&$220/242$\\ \hline
			\multicolumn{4}{c}{\mytorus\ model}  \\ \hline 
		$\Gamma$ & \aer{1.71}{0.09}{0.05}&\ser{1.60}{0.17}&\ser{1.79}{0.07}\\ 
		$N_{\textsc{pow}}$ (\tento{-3}) &\ser{6}{2}&\aer{1.4}{1.2}{0.7}&\aer{6.4}{1.0}{1.6}\\
		$\nh$ (\tento{22} \sqcm) &\ser{94}{7}&\ser{39}{8}&\ser{57}{6}\\
		$A_S=A_L$ &\aer{0.25}{0.22}{0.19}&\lsup{2.4}&\lsup{0.2}\\
		$f_s$ (\tento{-2})&\aer{0.11}{0.06}{0.04}&\aer{1.5}{1.6}{0.8}&\aer{5.4}{1.2}{0.7}\\
				$\lowlum$ (\tento{43} \lumcgs)&\ser{4.7}{1.0}&\ser{5}{1}&\ser{31}{5}\\
				$\highlum$ (\tento{43} \fluxcgs)&\ser{12}{1}& \ser{15}{3}&\ser{67}{3}\\
%		$kT_{\apec}$ (keV)&\aer{0.86}{0.17}{0.14}&&\\
%		$N_{\apec}$ (\tento{-6})&\aer{3.3}{2.1}{1.4}&&\\
		%		$\lowflux$ (\tento{-12} \fluxcgs)&&&\\
		%		$\highflux$ (\tento{-11} \fluxcgs)&&&\\
		$\rchisq$&$121/129$&$140/151$&$223/241$\\ \hline	
	\end{tabular}
	\end{center}
\end{table} 

\begin{table}
	\begin{center}
		%\scriptsize
		\caption{Properties of the neutral, narrow \fek\ emission line as found from the \zphabs\ model. For \tc, we fixed the energy at 6.4 keV to test the significance of the line in the \nus\ data. \label{tab:line}}
		\begin{tabular}{ l c c c} 
			\hline \hline
			& \ngc\ & \fc & \tc  \\ \hline 
			Rest-frame energy (keV) &\ser{6.31}{0.08}&\ser{6.45}{0.15}& 6.4(f)\\
			Flux (\tento{-6} photons \sqcm\ s$^{-1}$) &\ser{2.6}{1.6}&\ser{4}{3}& $<3$\\ 
			Equivalent width (eV) &\aer{110}{80}{60}&\ser{120}{100}& $<50$ \\
			\hline		
		\end{tabular}
	\end{center}
\end{table}

\section{Discussion and conclusions}\label{sec:discussion}
We have reported results based on \nus\ observations of the three absorbed radio galaxies \ngc, \fc\ and \tc, selected as possible CT candidates among the sample of \panessa. We derived constraints on their spectral parameters, first and foremost the absorbing column density $\nh$. None of these sources is found to be formally Compton-thick, i.e. $\nh<\expom{1.5}{24}$ \sqcm\ in all three. The most absorbed one is \ngc, for which we derived $\nh\simeq  \expom{1}{24}$ \sqcm. 

In all cases, we found no evidence for a strong Compton reflection component, neither in the approximation of reflection off a slab of infinite optical depth, nor assuming a more physical torus model.
Moreover, the \fek\ line at 6.4 keV is weak (see Table \ref{tab:line}).
These results are consistent with what is generally found in radio galaxies, which show weaker X-ray reprocessed features than Seyferts \cite[e.g.][]{wozniak1998,eracle2000,grandi2001}. 
In the past few years, \nus\ provided us interesting results on single radio sources. For example, the \nus\ spectra of the broad-line (i.e. type 1) radio galaxies 3C~382 and 3C~390.3 show a weak reflection component \cite[]{ballantyne2014,lohfink2015}, and Centaurus~A (which is optically a type 2) shows no evidence for reflection at all \cite[]{fuerst2016}. However, the \nus\ spectrum of Cygnus~A (also a type 2) is consistent with significant reflection from Compton-thick matter, although the absorption along the line of sight is Compton-thin \cite[]{reynolds2015}. The \nus\ spectrum of the radio-loud quasar 4C~74.26 is also consistent with ionized reflection \cite[]{lohfink2017}.
In general, there can be different explanations for the weak reflection from the accretion disc (which would be expected if the disc is optically thick). In particular, the X-ray corona cold be outflowing, so that its radiation is beamed away from the accretion disc and the surrounding material \cite[e.g.][]{belo1999,malzac2001}. This scenario could be promising for jetted sources, where the corona could be the base of the jet itself \cite[e.g.][]{lohfink20133c120,fabian2014}. Another possibility is that the inner region of the accretion disc is optically thin, like an advection-dominated accretion flow \cite[e.g.][]{adaf}. Moreover, if we interpret the absorption in our sources as being due to a pc-scale torus, the accompanying reflection component is small (if any). This might suggest that the torus has a small covering factor, consistently with the spectral modelling with \mytorus. The equivalent width of the \fek\ line alone provides a less model-dependent diagnostics. Following the detailed calculations by \cite{yaqoob5548}, the expected equivalent width (EW) of the line can be expressed as:
\begin{equation}
\mathrm{EW} \simeq 40 \left( \frac{f_c}{0.35} \right) \left( \frac{\nh}{\tentom{23} \; \mathrm{cm}^{-2} } \right) \; \mathrm{eV}
\end{equation}
where $f_c$ is the covering factor of the line-emitting material (spherically distributed), assuming solar Fe abundance and an incident power law with $\Gamma=1.65$. In the case of \ngc, for example, we derive a covering factor $f_c < 0.17$ from the constraints on the \fek\ line.

The apparent lack of reflection in the \nus\ spectrum of \tc\ is at odds with the results of \cite{fioretti2013}, who reported a reflection-dominated spectrum based on the 2007 \suz\ data. On the other hand, the absorbing column density is consistent between the two observations. We note that the 2--10 keV flux was found to be \expo{1.85}{-12} \fluxcgs\ in the 2007 \suz\ observation \cite[]{fioretti2013}, while we measured \expo{3.4}{-12} \fluxcgs\ in the 2017 \nus\ observation. This difference of a factor of almost 2, combined with the similar absorption, indicates an intrinsic luminosity variation, the source being possibly fainter and reflection-dominated during the 2007 \suz\ exposure. This behaviour might resemble that of the so-called `changing-look' CT AGNs, namely sources undergoing changes from Compton-thin to reflection-dominated states (and vice-versa) likely due to dramatic variations of the nucleus luminosity \cite[e.g.][and references therein]{matt2003}. On the other hand, we might have expected a stronger flux variation in \tc\ associated to this `change of look'. Further observations with \nus\ (which provides a much higher signal-to-noise above 10 keV compared with \suz) would be needed to confirm this puzzling behaviour.

Concerning the scattered fraction, we derived a value of 0.2-0.3 per cent for \ngc, around 2 per cent for \fc, and around 9 per cent for \tc. These values are reduced by a factor of 2 when assuming a more physical torus model. They are also roughly in agreement with previous estimates on \ngc\ \cite[0.55 per cent according to][]{eguchi2011} and \fc\ \cite[from the best-fitting parameters of][]{evans2008}, whereas for \tc\ \cite{fioretti2013} reported only an upper limit of 0.5 per cent. As we noted above, this discrepancy can be simply due to the contamination of the \nus\ spectrum by the diffuse emission at energies lower than 5 keV. The scattered fraction $f_s$ could be related with the geometry of the torus. Indeed, a small value of $f_s$ (say below 0.5 per cent) might indicate a ``hidden'' or ``buried'' AGN, surrounded by a very geometrically thick torus \cite[]{ueda2007,winter2009}. However, this is at odds with the lack of strong reflection by the torus itself. It would be probably more reasonable to infer a small amount of scattering material. Furthermore, other interpretations are possible. For example, rather than Compton-thin scattering, the ``soft excess'' could be due to photoionized mission \cite[]{gb2007}. To test this scenario, high-resolution soft X-ray observations would be needed.

From the 2--10 keV luminosities that we derived, we can estimate the Eddington ratio of the sources. We assume for the luminosity the values found with the \mytorus\ model, which takes into account Compton scattering. \ngc\ has an estimated black hole mass of around \expo{4}{8} solar masses \cite[from the stellar velocity dispersion;][]{bettoni2003}, and \tc\ has a similar black hole mass of around \expo{3}{8} solar masses \cite[from the host bulge magnitude;][]{marchesini2004}. We are not aware of estimates of the black hole mass of \fc. Using the 2--10 keV bolometric corrections of \cite{marconi2004}, we estimate the bolometric luminosities to be $\sim \expom{1.3}{45}$ \lumcgs\ for \ngc\ and $\sim \expom{1.6}{46}$ \lumcgs\ for \tc. Given the black hole masses, the Eddington ratios are of the order of 0.02 for \ngc\ and 0.37 for \tc, respectively.
3C~452 is thus in a high-luminosity, strongly accreting state.  
\ngc\ is an order of magnitude less luminous and weaker in terms of accretion rate, but can still be considered a high-luminosity and highly accreting AGN \cite[e.g.][]{panessa2006,kollmeier2006}.

\subsection{CT radio galaxies}\label{subsec:CT}
According to our results, the constraints on the column density indicate that the three sources that we considered are absorbed but not CT, consistently with the lack of strong reflection. We now examine the radio galaxies that have been reported as CT in the literature, and already mentioned in Sect. \ref{sec:intro}.

\subsubsection{Compact sources}\label{subsubsec:compact}
Mrk~668 (OQ+208) is a dust-obscured galaxy \cite[DOG;][]{hwang2013_a,hwang2013_b} hosting a compact, Gigahertz-Peaked Spectrum (GPS) radio source. This AGN has been reported as the first CT broad-line radio galaxy, from the analysis of \xmm\ data \cite[]{guainazzi2004}. Indeed, the hard X-ray spectrum was found to be flat and showed a prominent neutral \fek\ line with equivalent width of 600 eV, suggesting a reflection-dominated spectrum, while the absorbing column density was found to be larger than \expo{9}{23} \sqcm\ \cite[]{guainazzi2004}. 
%This source is also a dust-obscured galaxy \cite[]{hwang2013_a,hwang2013_b}.

PKS~1607+26 is also a GPS, labelled as a compact symmetric object (CSO). From \xmm\ data, \cite{tengstrand2009} suggested that this source could be CT ($\nh > \expom{6}{23}$ \sqcm). 
From \chandra\ data, \cite{siemiginowska2016} later showed that the source is accompanied by a close, secondary X-ray source of uncertain nature, that \xmm\ is not capable to resolve. 
Moreover, \cite{siemiginowska2016} found that the \chandra\ spectrum is consistent with being essentially unobscured. This might suggest a changing-look scenario, but the presence of the secondary source complicates the interpretation and different scenarios are possible \cite[see][]{siemiginowska2016}.

TXS~2021+614 is another GPS/CSO. The \chandra\ spectrum is found to be flat ($\Gamma\simeq0.8$), and is consistent with absorption by $\nh \gtrsim \expom{9.5}{23}$ \sqcm\ plus reflection \cite[]{siemiginowska2016}. However, also given the lack of a significant \fek\ line, an alternative explanation based on non-thermal X-ray lobe emission cannot be ruled out \cite[]{siemiginowska2016}.

Incidentally, we note that GPS/CSOs are young sources which could represent an early stage of the evolution of radio galaxies. Their X-ray absorption is not necessarily due to a torus, especially if the X-ray emission is produced by jets and/or lobes, which would rather be absorbed at galactic scales \cite[e.g.][see also Sect. \ref{subsec:absorber}]{ostorero2016,siemiginowska2016}. 

\subsubsection{Extended sources}\label{subsubsec:extended}
Apart from the three compact radio galaxies mentioned above, the situation is even less clear for extended radio galaxies, and only a few of them have been proposed as CT candidates. Among the local objects, these are 3C~284, 3C~223 and 3C~321, in addition to the three sources examined in this paper.

3C~284 is a FRII radio source, hosted by a disturbed elliptical with dust lanes \cite[]{floyd2008}. \cite{croston2004} reported an X-ray absorbing column density $\nh=\expom{\aerm{2.6}{1.4}{1.0}}{23}$ \sqcm\ from the analysis of \xmm\ data. However, \cite{hardcastle2006} reported $\nh>\expom{1.2}{24}$ from the same data set, ascribing the discrepancy to changes in the \xmm\ calibration files. A similar column density ($\nh=\expom{\aerm{2.0}{1.4}{0.7}}{24}$ \sqcm) was reported by \cite{corral2014}, from the same \xmm\ data. 

3C~223 is another FRII radio galaxy, for which \cite{croston2004} reported a column density smaller than \expo{1}{23} \sqcm\ based on \xmm\ data. \cite{hardcastle2006} found $\nh=\expom{\aerm{5.7}{7.8}{3.6}}{23}$ \sqcm\ from their re-analysis of the same data. \cite{corral2014} instead suggested that the \xmm\ spectrum is reflection-dominated, estimating a column density larger than \tento{24} \sqcm. Finally, from the same data set \cite{lamassa2014} suggested that the source is surrounded by a globally Compton-thin medium ($\nh \sim \tentom{23}$ \sqcm), giving rise to the reflection component, while the line-of-sight column density is Compton-thick ($\nh > \expom{1.67}{24}$ \sqcm). This would be a rather unique case among absorbed AGNs. 

3C~321 is a FRII radio galaxy with a peculiar radio morphology, and in the process of merging with a companion galaxy, which might also host an active nucleus \cite[]{evans2008}. Moreover, a dust lane is seen at kpc scales \cite[e.g.][]{martel1999}. From the analysis of \chandra\ data, \cite{evans2008} found the nucleus of 3C~321 to be obscured by a column density $\sim \tentom{24}$ \sqcm. However, at least part of this obscuration could be due to the complex galaxy environment.

Finally, \cite{wilkes2013} reported eight CT candidates among the high-redshift ($1<z<2$) sources of the 3CRR catalogue \cite[]{laing1983}, from \chandra\ data.
%These are 3C 13/68.2/266/324/356/368/437 and 4C 13.66. 
Given the low number of counts, this suggestion is not based on spectral fits, but mostly on the X-ray hardness ratio, on the X-ray to radio luminosity ratio, and on the [\oiii]$\lambda 5007$ emission line. The estimated column densities for these eight candidates are $\sim \expom{2}{24}$ \sqcm, but the faintness of the sources makes this result quite uncertain \cite[see the discussion in][]{wilkes2013}.

Taken together, these findings indicate that a handful of radio galaxies could be consistent with being heavily absorbed. On the other hand, the uncertainties are large, also because of the lack of high-quality data at high energies. As a result, we currently lack clear-cut evidences for formally CT radio galaxies, i.e. with $\nh$ well constrained to be larger than \expo{1.5}{24} \sqcm. Of course this threshold is somewhat arbitrary, but it does have a physical meaning, namely the absorbing material becomes optically thick to Compton scattering.
In any case, in radio galaxies we seem to have an indication of rare (or even lacking) heavy absorption, i.e. exceeding a few times \tento{24} \sqcm, compared with radio-quiet Seyferts (see also \panessa). 

Apart from the few single sources discussed above, from \panessa\ and the present work we have no evidence for heavily absorbed sources among the hard X-ray selected sample of \cite{bassani2016}. 
An obvious selection bias against CT sources is absorption itself, which can produce a strong reduction of flux. Even at hard X-rays, a CT column density implies significant Compton downscattering to energies where photoelectric absorption dominates \cite[e.g.][]{matt1999}. Therefore, even hard X-ray surveys are not totally immune from the absorption bias \cite[e.g.][]{malizia2009,burlon2011,malizia2012}.
%However, the observed fraction of CT AGNs among radio-quiet, hard X-ray selected sources is of at least 5--7 per cent \cite[e.g.][]{malizia2009,malizia2012}. Then, assuming similar absorption properties, we might have expected to observe at least 3-4 CT sources among the 64 radio galaxies of \cite{bassani2016}, while do not have strong evidence for any. 
However, even if the numbers are not large enough to draw definitive conclusions, we have a tentative indication that only a few radio galaxies (if any) are CT. This might hint for a difference between the absorption properties of radio galaxies and radio-quiet Seyferts, which deserves further investigation.

\subsection{What is the nature of the absorber?}\label{subsec:absorber}
The neutral absorption seen in radio galaxies, just like radio-quiet Seyferts, is generally ascribed to a putative pc-scale torus \cite[e.g.][]{urry&padovani,hardcastle2009}. The presence of a molecular, dusty torus in Seyferts is also widely invoked to explain the infrared (IR) emission, which is often interpreted as thermal emission from hot dust \cite[e.g.][]{pier1993}. In recent years, optical/IR interferometry permitted to resolve the pc-scale environment of AGNs and directly observe the torus \cite[e.g.][and references therein]{burtscher2013}. The basic tenet of a torus-dominated IR emission has been challenged by the observation of strong mid-IR emission originating from the polar region at 10--100 pc-scales, rather than the torus \cite[e.g.][]{honig2013,asmus2016}. On the other hand, the near-IR emission is likely to originate from the equatorial plane \cite[e.g.][]{honig2013}. A two-component disc+outflow scenario is thus emerging, where dust clouds are accreted in the disc plane and are in part lifted up by radiation pressure, forming an outflow cone \cite[]{honig2017}. Future, multiwavelength observations on radio galaxies will be needed to gain further insight on the relation between the mechanisms of accretion (disc/torus) and ejection (jet/outflows). Interestingly, \ngc, namely the most absorbed source reported in \panessa\ and discussed here, does not show evidence for any mid-IR emission from the nucleus, being dominated by star formation at much larger scales \cite[]{asmus2014,duah2016}. This points to a relatively weak AGN, in agreement with the low Eddington ratio (0.02) that we have estimated. Moreover, given the peculiar morphology of this source, we may also speculate that the X-ray absorption is (at least partly) due to a large-scale structure rather than to a pc-scale torus.

In general, it is possible that at least part of the absorption is due to different material than the torus, eventually located further away and even at galactic scales.  For example, the presence of dust lanes is ubiquitous in Seyferts \cite[e.g.][]{malkan1998} and it is likely related with Compton-thin X-ray obscuration \cite[]{matt2000_dust,gmp2005}.
Furthermore, 21 cm \hi\ absorption is commonly observed in radio AGNs and it can trace rotating discs, outflows or more complex morphologies \cite[e.g.][]{morganti2001,gereb2015}.
Recently, \cite{glowacki2017} and \cite{moss2017} reported a correlation between X-ray absorption and 21 cm absorption in a sample of obscured radio AGNs. Such a correlation was also found by \cite{ostorero2010,ostorero2016} for GPS sources, suggesting that the X-ray obscuration might be caused by gas at large scales \cite[][]{siemiginowska2016}. It remains to be seen whether the 21 cm absorption is due to a pc-scale torus, or to a galactic inflow at a few tens of pc-scale, or to a kpc-wide galactic structure \cite[]{moss2017}. Interestingly, old, extended radio sources seem to show a lower fraction of 21 cm \hi\ detection than young, compact ones \cite[][]{chandola2013}. 
Then, if the \hi\ and X-ray absorption are truly related, we expect to observe weaker X-ray absorption in extended radio sources.
Among the sources discussed in this paper and in Sect. \ref{subsubsec:extended}, \hi\ absorption has been detected in \ngc\ \cite[][see also Sect. \ref{sec:intro}]{emonts2008}, in \tc\ \cite[]{gupta2006,chandola2013} and in 3C~321 \cite[]{chandola2012}. The column density is estimated to be $\nhi \simeq \expom{5}{21}$ \sqcm\ for \ngc\ \cite[]{emonts2008}, $\nhi \simeq \expom{6.4}{20}$ \sqcm\ for \tc\ \cite[]{gupta2006} and $\nhi \simeq \expom{9.2}{21}$ \sqcm\ for 3C~321 \cite[]{chandola2012}. $\nhi$ seems thus to be much smaller than the column density measured in X-rays ($\nh \sim \tentom{23-24}$ \sqcm). 
However, the inferred $\nhi$ is generally considered a lower limit,
%may be in error by up to 3 orders of magnitude
because of a degeneracy with the (unknown) spin temperature and covering fraction \cite[]{gupta2006,gereb2015}.
%In principle, a measure of $\nhi$ can also come from Lyman~$\alpha$ absorption, but at larger redshifts. 
Owing to such uncertainties, we cannot rule out a significant contribution to the X-ray column density by atomic hydrogen seen at radio frequencies, and that could reside at distances larger than the pc-scale torus.

\section{Summary}\label{sec:summary}
We have studied the X-ray spectra of three CT candidates (\ngc, \fc\ and \tc) among the hard X-ray selected radio galaxies investigated in \panessa, making use of \nus, \xmm, \swift/BAT and \integral\ data. 
Our conclusions can be summarized as follows:\\
- We do not find formally CT sources. \ngc\ is the only source consistent with a column density of around \expo{1}{24} \sqcm. No source shows evidences for a strong reflection component. Comparing with a past \suz\ observation, it is possible that \tc\ switched from a reflection-dominated to a Compton-thin state over the course of 10 years.\\
- The lack of strong evidences for CT hard X-ray selected radio galaxies, also coupled with residual uncertainties on the real nature of the few CT radio galaxies reported in the literature, is quite puzzling. 
Although we cannot rule out the effect of a selection bias, 
this result could hint for a discrepancy between the average absorption properties of radio-loud and radio-quiet AGNs.\\
- The origin of the X-ray absorbing medium is not obvious. A significant role could be played by material different from the classical pc-scale torus, such as that traced by 21 cm \hi\ absorption, which can be located much farther away (e.g. at galactic scales for \ngc). 
\section*{Acknowledgements}
We thank the referee for helpful comments that improved the paper.

 We acknowledge the use of public data from the \nus, \xmm, \swift\ and \integral\ data archives. 
This research has made use of data, software and/or web tools obtained from NASA's High Energy Astrophysics Science Archive Research Center (HEASARC), a service of Goddard Space Flight Center and the Smithsonian Astrophysical Observatory, and of  the \nus\ Data Analysis Software jointly developed by the ASI Space Science Data Center (SSDC, Italy) and the California Institute of
Technology (USA).
We acknowledge financial support from ASI under contract ASI/INAF 2013-025-R01.

\bibliographystyle{mnras}
\bibliography{mybib.bib}

\bsp	% typesetting comment
\label{lastpage}
\end{document}